\documentstyle[sprocl,epsf]{article}

\def\NPB{{\em Nucl. Phys.} B}
\def\PLB{{\em Phys. Lett.}  B}

\def\PRD{{\em Phys. Rev.} D}

\def\be{\begin{equation}}
\def\ee{\end{equation}}
\def\bea{\begin{eqnarray}}
\def\eea{\end{eqnarray}}
\def\Vec#1{\mbox{\boldmath $#1$}}

\def\itmb{\begin{itemize}}
\def\itme{\end{itemize}}
\def\enmb{\begin{enumerate}}
\def\enme{\end{enumerate}}
\def\eqnb{\begin{equation}}
\def\eqne{\end{equation}}
\def\eqab{\begin{eqnarray}}
\def\eqae{\end{eqnarray}}
\def\dis{\displaystyle}

\begin{document}

\title{Numerical study of the Kugo-Ojima criterion and\\
 the Gribov problem in the Landau gauge}

\author{Hideo NAKAJIMA}

\address{Department of Information science, Utsunomiya University, 
Utsunomiya,\\ 321-8585 Japan\\E-mail: nakajima@is.utsunomiya-u.ac.jp}

\author{Sadataka FURUI }

\address{School of Science and Engineering, Teikyo University, 
Utsunomiya,\\ 320-8551, Japan\\E-mail: furui@dream.ics.teikyo-u.ac.jp} 

\author{Azusa YAMAGUCHI}

\address{Particle Physics Lab. Department of Physics, Ochanomizu University,
Utsunomiya,\\ 112-8610, Japan\\E-mail: azusa@sokrates.phys.ocha.ac.jp}


\maketitle\abstracts{The Kugo-Ojima color confinement criterion,
which is based on the BRST symmetry of the continuum QCD is
numerically tested by the lattice Landau gauge simulation.
We first discuss the Gribov copy problem and the BRST symmetry
on the lattice. The lattice Landau gauge can be formulated with
options of the gauge field definition, $U$(link)-linear type or
$\log U$ type.
The Kugo-Ojima parameter $u^a_b$ which is expected to be
 $-1\delta^a_b$ in the
continuum theory is found
to be $-0.7\delta^a_b$ in the strong
coupling region, and the magnitude is a little less in the weak coupling
region in $\log U$ type simulation. Those values are weakened
 even further in $U$-linear type.
The horizon function defined by Zwanziger is evaluated in
 both types of gauge field and compared. 
The horizon function in the $\log U$ version is larger than the other,
but in the weak coupling region, the expectation 
value of the horizon function is suggested to be zero or negative.}

\section{Introduction}
There are essentially two aspects in the manifestation of color confinement in
the Landau gauge QCD. One aspect is the linear potential between quarks, 
which was conjectured by Gribov as a consequence of an
enhancement of the singularity of the ghost propagator\cite{Gv} due to the 
restriction of the gauge field $A$ on the transverse plane.  
Another aspect is the absence of free single colored particle state in the 
asymptotic Hilbert space, which culminates in the Kugo and Ojima color 
confinement criterion based  on the BRST(Becchi-Rouet-Stora-Tyutin) symmetry:
i.e. in the Landau gauge, a coefficient in the two-point function produced by 
the ghost, the antighost and the gauge field becomes $-\delta_a^b$, where 
$a$ and $b$ specify the color in the adjoint representation. Analytical 
calculation of this value is extremely difficult and so far no verification 
has been performed.
In 1994,  Zwanziger developed a lattice QCD theory for Gribov ambiguity.
He claimed that, if the restriction to the fundamental modular region is
 achieved, the gluon propagator at 0 momentum should vanish in the 
continuum limit.

As the matter of first principle, it is still unclear
whether the bases of these two theories are consistent or not, that is,
'Does the Kugo-Ojima theory properly resolve the problem of
Gribov copy?' and/or 'Is the Zwanziger theory satisfactorily
well defined in the
continuum limit?' Considering this fundamental problem
on one side, we make numerical verifications on conjectures
of these theories on the other.
In the lattice QCD test, we address the following 
problems:  whether the gluon propagator is infrared finite, how singular the ghost propagator is in the infrared region,
 whether the Kugo-Ojima color confinement criterion\cite{KO} is satisfied,
and how close to zero from negative the Zwanziger horizon function is.

\section{The Gribov problem and the lattice simulation of the Landau gauge QCD
}
\subsection{The path integral formulation of the gauge fixed theory in the
presence of the Gribov copy (Fujikawa, Hirschfeld)}

 First we give a brief review of the path integral formulation of the
gauge fixed theory by Fujikawa\cite{Fuj} and Hirschfeld\cite{Hir}. 
They discussed a possible
situation which may give a way out of the gauge fixing degeneracy problem
(the Gribov problem).
In the following, all indices representing multi-degree of freedom are
suppressed.
Let us define a gauge unfixed partition function as,
\begin{equation}
Z=\int dU e^{-\beta S(U)}.
\end{equation}
In derivation of the Faddeev-Popov(Faddeev-Popov) formula of the gauge $f(U)=0$,
one considers the determinant function $\Delta (U^g)$,
\[
\Delta (U^g)=\det \left(\displaystyle {\partial f(U^g)\over \partial U^g}
\displaystyle {\partial U^g\over \partial g}\right )
=\det \left(\displaystyle {\partial f(U^g)\over \partial A^g}
\displaystyle {\partial A^g\over \partial g}\right )
\] 
where $A_\mu=A_\mu(U)$, and one finds that $\Delta(U^g)$ is indeed
a function of $U^g$, and looks at the integral on the gauge orbit,
$U^g$,
\[
N(U)\equiv\int dg \Delta (U^g) \delta(f(U^g)).
\]
Obviously $N(U)$ is an orbit function, that is, $N(U^g)=N(U)$.
At the intersection points of the gauge orbit $U^g$ with the surface $f(U)=0$,
i.e., $g=g_i(U)$, Gribov's copies, the above delta function is transcribed
to give
\begin{equation}
N(U)=\int dg \sum_i
\displaystyle{\Delta (U^g) \over |\Delta (U^g)|}\delta(g-g_i(U))
=\int dg \sum_i sign(\Delta (U^g))\delta(g-g_i(U)).
\label{EXX1}
\end{equation}
These delta functions contribute $0$ or $\pm 1$, and thus, if the orbit
function $N(U)$ is non vanishing over all orbits, $N(U)\ne 0$, then in
use of the identity
\[
1=\displaystyle{1\over N(U)}\int dg \Delta (U^g) \delta(f(U^g)),
\]
the standard FP procedure applies, and one factors out the
gauge volume and
obtains the formula
for expectation values of functions, $F(U)$,
\begin{equation}
\langle F(U)\rangle|_{f\ gauge}=
\displaystyle{\int dU \Delta (U)\delta (f(U))F(U)e^{-\beta S(U)}/N(U)
\over
\int dU \Delta (U)\delta (f(U))e^{-\beta S(U)}/N(U)},
\label {EX1}
\end{equation}
provided $F(U)$ is gauge invariant. In the case when
$F(U)$ is gauge non-invariant, $F(U)$ in (\ref{EX1}) 
should be formally replaced with a gauge invariant function,
\eqnb
\tilde F(U)=\dis{\int dg F(U^g)\over \int dg},
\label{AVF}
\eqne
if one really wishes to obtain the Boltzmann average of the
gauge non-invariant function $F(U)$.
The formula (\ref{EX1}) allows the BRST formulation as in the
following, and if 
$N(U)$ is particularly a constant, then the standard Lagrangian of the
continuum theory is justified. The formula (\ref{EX1}) also
derives a natural lattice simulation algorithm of the gauge fixed theory.
The treatment of the gauge fixed theory here is, however,
somewhat too formal for its practical use, particularly with respect to
gauge non-invariant function in the presence of Gribov copies.
We discuss this point in the following section.

\subsection{The simulation algorithm for the gauge fixed
 theory (Mandula-Ogilvie) and the Gribov problem}
Multiplying by the gauge volumes, the denominator and the numerator of the
formula (\ref{EX1}), respectively, one recovers the $Z$ in the denominator,
and has in the numerator,
\begin{equation}
\int dU \int dg \Delta (U^g)\delta (f(U^g))F(U^g)e^{-\beta S(U)}
/N(U).
\end{equation}
As in (\ref{EXX1}), one obtains the numerator as
\begin{equation}
\int dU \sum_i sign(\Delta(U^{g_i}))F(U^{g_i})e^{-\beta S(U)}/N(U)
\end{equation}
where $U^{g_i}$ is the i-th Gribov copy on the orbit $U^g$. This gives us
the algorithm in the simulation\cite{MO}
\begin{equation}
\langle F(U)\rangle|_{f\ gauge}
=\displaystyle{1\over Z}\int dU \sum_i \displaystyle{sign(\Delta(U^{g_i}))F(U^{g_i})\over N(U)}e^{-\beta S(U)}
=\langle \bar F(U)\rangle
\label {MDOEQ1}
\end{equation}
where the last averaging $\langle\rangle$ is that of simulation,
i.e., that with respect to the Boltzmann weight $e^{-\beta S}$, and
$\bar F(U)$ is a sign-weighted average of $F(U^{g_i})$ on the
gauge orbit $U^g$. Readers should have already noticed that (\ref{MDOEQ1})
gives trivial results when $F(U)$ is gauge invariant, that is,
we do not need any gauge transformation at all.
Necessity of the function $\tilde F(U)$ is rather
impractical as well, and
therefore we
start from the formula (\ref{EX1}) with a gauge non-invariant
function $F(U)$
for work purpose. Then the result (\ref{MDOEQ1}) is highly
nontrivial, and $\bar F(U)$ can be viewed as being defined as a
new gauge invariant function from $F(U)$ and $f(U)$.
The sign-weighted average is impractical and a simple
average\cite{Bny} may be similarly considered as such.
If the averaging has, however, a large fluctuation
this formula would rather be interpreted as representing Gribov noise
apart from the original function $F(U)$,
although (\ref{MDOEQ1}) could be considered as giving a sort of gauge invariant
function out of $F(U)$ and $f(U)$.
A modification of the gauge such that one chooses a unique copy
among others on the orbit is favored, that is, {\bf a new gauge without 
Gribov copy}, and in that case, the above formula is
useful in practice, and this procedure may naturally change the result
in the above formula ({\ref{MDOEQ1}}) without Gribov noise.

\subsection{The BRST formulation and the Gribov problem}
The obvious standard FP formula allows BRST formulation,
\begin{equation}
\langle F(U)\rangle|_{f\ gauge}=\displaystyle{1\over Z'}
\int d\mu \exp( \delta \int \bar c f(U))F(U)e^{-\beta S(U)}/N(U),
\label {EX2}
\end{equation}
\begin{equation}
Z'=\int d\mu \exp( \delta \int \bar c f(U))e^{-\beta S(U)}/N(U),
\label {EX3}
\end{equation}
where $\delta$ stands for  BRST transformation,
\begin{equation}
\delta \bar c=i B,\ \ \delta A_\mu(U)=D_\mu(U)c,\ \ 
\delta c=-cc,\ \ \delta B=0,
\label {EX4}
\end{equation}
and the measure is defined as
$d\mu=dUdBdcd\bar c$,
where $dB$ integration is performed on ${\Vec R}$ and $dcd\bar c$ is
suitablly defined differentiations with respect to Grassmann numbers, $c$ and
$\bar c$
, such that
\begin{equation}
\int dcd\bar c\ e^{\bar c M c}=\det M,
\label {EX6}
\end{equation}
\begin{equation}
\int dcd\bar c\ c_x \bar c_y e^{\bar c M c}=(\det M)(-M^{-1})_{xy}.
\label {EX7}
\end{equation}

It was shown by Neuberger that if the gauge fixing function $f(U)$ is a smooth
function of compact variables $U$, then the expectation value
of gauge invariant function $F(U)$ becomes an indefinite form\cite{Neu},
\begin{equation}
\langle F(U)\rangle|_{f\ gauge}=\displaystyle{0\over 0},
\end{equation}
which implies that all Gribov copies contribute to give total cancellation,
in other words, the assumption that the above $N(U)$ is non vanishing on all
orbits, does not hold, and the formula (\ref{EX1}) is totally meaningless.\par
An essential point of his argument is as follows.
Let us consider a general expression
\begin{equation}
Z'(t)=\int d\mu \exp(t \delta \int \bar c f(U))G(U)
\label {EX8}
\end{equation}
with a gauge invariant function $G(U)$ as the Boltzmann weight. Then
one finds that
\begin{equation}
\displaystyle{dZ'\over dt}(t)=\int d\mu (\delta \int \bar c f(U))
\exp(t \delta \int \bar c f(U))G(U)
\label {EX9}
\end{equation}
can be written from nilpotency of $\delta$, i.e., $\delta^2=0$, and
from $\delta G(U)=0$ as,
\begin{equation}
\displaystyle{dZ'\over dt}(t)=\int d\mu \delta \left((\int \bar c f(U))
\exp(t \delta \int \bar c f(U))G(U)\right).
\label {EX10}
\end{equation}
If $W(U,B,c,\bar c)$ is an analytic function of the compact variables $U$, 
then one can show that
\begin{equation}
\int d\mu \delta W(U,B,c,\bar c)=0.
\label {EX11}
\end{equation}
Thus it follows that
Together with $Z'(0)=\int \cdots d\bar c G(U)=0$, one finds that
$Z'(t)=0$, so $Z'(1)=0$. This concludes total cancellation of
the Gribov copies of gauges given by analytic gauge functions of $U$.
Thus one is forced to consider non-analytic gauge functions of $U$ as
desired gauge functions. As a one dimensional $U(1)$ toy example
which avoids (\ref{EX11}),
one can consider  $U=e^{iA}$, $A={\rm Im} \log U$, where $A$ is not continuous
at $U=-1$, and with definition $\delta A=c$, one finds that
\begin{equation}
\int dUdBdcd\bar c\ \delta( e^{-B^2}\bar c A)=
\int_{-\pi}^{\pi} dA\int _{-\infty}^{\infty}dBdcd\bar c\ e^{-B^2}c\bar c\ne 0.
\label {EX12}
\end{equation}
Although this problem is still open in the lattice BRST formulation,
use of non compact variables in gauge fixing functions may be helpful.

\subsection{The Landau gauge and the Gribov problem}
Now we focus on the Landau gauge in $SU(3)$ lattice QCD, that is,
$f(U)=\partial_\mu A_{\mu}(U),$
where there are some options of definition $A_{\mu}(U)$ as 
\begin{enumerate}
\item $U$-linear one\cite{Zw};
$A_{x,\mu}=\displaystyle{1\over 2}(U_{x,\mu}-U_{x,\mu}^{\dag})|_{traceless\ part},$
\item
use of exponential map\cite{NF}.
$U_{x,\mu}=\exp{A_{x,\mu}},\ \ \ A_{x,\mu}^{\dag}=-A_{x,\mu},$
where absolute values of all eigenvalues of $A_{x,\mu}$ do not exceed $4\pi/3$.
\end{enumerate}
In the latter definition, $A_{x,\mu}(U)$ is not analytic with
respect to compact variable $U_{x,\mu}$ contrary to the former one.
In both cases the Landau gauge, $\partial A^g=0$, can be
 characterized\cite{MN,SF} in use
of optimizing functions $F_U(g)$ of $g$, such that $\delta F_U(g)=0$ for
any $\delta g$.
\begin{enumerate}
\item $U$-linear definition;
$F_U(g)={2N\over dV(N^2-1)}\sum_{x,\mu}\left (1- {1\over N}{\rm Re}{\rm tr}U^g_{x,\mu}\right).$
\item use of exponential map;
$F_U(g)=||A^g||^2={1\over dV(N^2-1)}\sum_{x,\mu}{\rm tr}
 \left({{A^g}_{x,\mu}}^{\dag}A^g_{x,\mu}\right)$,
where $N$ is the number of colors and $d$ is the dimension. 
It is noteworthy that 
$||A^g||^2$ is a continuous function of compact
variables $U^g$ in spite of non analytic property of $A_{x,\mu}(U^g)$.
\end{enumerate}

In both options of the gauge field definition, the variation
of the optimizing function, $F_U(g)$, under infinitesimal gauge transformation
$g^{-1}\delta g=\epsilon$, reads as
\eqnb
\Delta F_U(g)=-2\langle \partial A^g|\epsilon\rangle+
\langle \epsilon|-\partial { D(A^g)}|\epsilon\rangle+\cdots,
\label{NORM1}
\eqne
where $D\mu(A)$ denotes the covariant derivative in each definition;
\enmb
\item
in $U$-linear version,
\eqnb
D_{\mu}(U_{x,\mu})\phi=\dis{1\over 2}\left\{ \dis{U_{x,\mu}+U_{x,\mu}^\dag\over 2},\partial_{\mu}\phi\right\}+[A_{x,\mu},\bar \phi],
\eqne
where
\eqnb
\partial_\mu \phi=\phi(x+\mu)-\phi(x),\ \ \ \bar \phi=\dis{1\over 2}\left(\phi(x+\mu)+\phi(x)\right),
\eqne
\item in $A=\log U$ version,
\eqnb
D_{\mu}(A_{x,\mu})\phi=S({\cal A}_{x,\mu})\partial_\mu \phi
+[A_{x,\mu},\bar \phi],
\eqne
where ${\cal A}_{x,\mu}=adj_{A_{x,\mu}}=[A_{x,\mu},\cdot]$, and 
\eqnb
S(x)=\dis{{x/2\over {\rm th}(x/2)}}.
\eqne
\enme

Gribov copy is generic phenomenon in both definitions as well as in the
continuum\cite{MN}, there exist
a lot of local minima of $F_U(g)$ along the gauge orbit $U^g$.
Thus the naive Landau gauge loses its solid basis
both in the theoretical and in the simulation view points for
examination of gauge non-invariant quantities such as
gluon propagator, ghost propagator, etc. 
\subsection{Sophisticated Landau gauge}
Zwanziger devised various regions of the transverse plane of $A^g$, i.e.,
$\partial A^g=0$, depending on properties of a point $U^g$ on the plane.
For example, one defines {\bf Gribov region} $\Omega$ as
\begin{equation}
\Omega=\{A|-\partial { D(A)}\ge 0\ ,\ \partial A=0\}\ \ ,
\label{GRIBOVR}
\end{equation}
where $D_\mu(A)$ denotes the covariant derivative, and $-\partial D\ge 0$ 
implies that the Faddeev-Popov operator $-\partial D$ is positive definite.
A point on the Gribov region is a local minimum of $F_U(g)$, but it
is known that some points on the Gribov region can be gauge copies of each
other. Thus one defines the {\bf fundamental modular region} $\Lambda$ as
the absolute minimum along the gauge orbits.
\begin{equation}
\Lambda=\{A|\|A\|^2={\rm Min}_g\|A^g\|^2\}, \qquad
\Lambda\subset \Omega\ \ .
\label{FNDMD}
\end{equation}
Then one defines a corresponding region 
${\cal U}_\Lambda$ of configuration $U$, as
\[
{\cal U}_\Lambda=
\{U|A(U)^T\in \Lambda\}
\]
\begin{equation}
(A^T\ {\rm denotes\ transverse\ component\ of\ A(U)})
\label{UFNDMD}
\end{equation}
Putting an indicator function $\theta_\Lambda$ of the set ${\cal U}_\Lambda$ as
\[
\theta_\Lambda(U)=1\ {\rm if}\ U\in {\cal U}_\Lambda,\ \ 
\theta_\Lambda(U)=0\ {\rm if}\ U\notin {\cal U}_\Lambda,
\]
and putting $\bar \theta_\Lambda(U)=1-\theta_\Lambda(U)$, one can define
a corresponding gauge function as
\begin{equation}
f_{\Lambda}(U)=\partial A\cdot \theta_\Lambda(U)+\bar \theta_\Lambda(U)
\label{UGFNCT}
\end{equation}
The arguments in the preceding subsections formally applies 
for a corresponding 
gauge function containing non-analytic Heaviside function. The gauge-fixing algorithm in the simulation is required
to attain the {\bf absolute minimum} of the $F_U(g)$ along the gauge orbit.
But the global minimization is difficult in general and 
developing the efficient algorithm is still an open problem\cite{YN}.


\section{The Kugo-Ojima confinement criterion and the Gribov-Zwanziger's theory}
\subsection{Kugo-Ojima's theory}

A sufficient condition of the color confinement given 
by Kugo and Ojima\cite{KO} is that
 $u^{ab}$ defined by the two-point
function of  the FP  ghost fields, $c(x),\bar c(y)$,
and $A_\nu(y)$,
\begin{equation}
\int e^{ip(x-y)}\langle 0|T D_\mu c^a(x)g(A_\nu\times \bar c)^b(y)|0\rangle d^4x=
(g_{\mu\nu}-{p_\mu p_\nu\over p^2})u^{ab}(p^2)
\label{eq}
\end{equation}
satisfies $u^{ab}(0)=-\delta^{ab}$.

Brief survey of Kugo-Ojima's argument
that $u^a_b=-\delta^a_b$ is a sufficient condition
of the colour confinement is the following.

\begin{enumerate}
\item The BRS transformation $\delta_B$ is given by replacing gauge
transformation
parameters $\theta^a$ as
$\theta^a(x) \to\ \lambda c^a(x)$, where $\lambda$ is an imaginary
Grassmann number and $c^a(x)$ is the ghost field.
Putting $C(x)=gc^a(x)\Lambda_a$ where $\Lambda_a$ is antihermitian such
that $[\Lambda_a,\Lambda_b]=f_{abc}\Lambda_c$
and $\delta_B=\lambda {\bf{\delta}}_B$, and antihermitian
matrix $A_\mu(x)=gA^a_\mu(x)\Lambda_a$, the BRS transformation reads
\[
{\bf{\delta}}_B
\phi=-C\phi,\ \ \ {\bf{\delta}}_B
A_\mu=D_\mu(A)C=\partial_\mu C+[A_\mu,C].
\]
The nilpotency requirement of $\delta_B$ derives,
\[
{\bf{\delta}}_B
C=-C^2.
\]
For each $c^a(x)$, one introduces an anti-ghost $\bar c^a(x)$, and
similarly the matrix $\bar C$, and
defines,
\[
{\bf{\delta}}_B
\bar C=iB,
\]
then the nilpotency gives
$
{\bf{\delta}}_B B=0
$. Then the gauge-fixing and Faddeev-Popov Lagrangian can be written
as
\begin{eqnarray}
\cal{L}_{GF+FP}&=&-i{\bf{\delta}}_B\large [\bar c^a\large(
 \partial^{\mu}A^a_{\mu}+\displaystyle{1\over 2}\alpha B^a
\large)\large]\nonumber\\
&=&B^a 
\partial^{\mu}
A^a_{\mu}+\displaystyle{1\over 2}\alpha B^a B^a
+i\bar c^a \partial^{\mu}D_{\mu}c^a.
\end{eqnarray}
Together with the original gauge invariant Lagrangian,
the BRS invariance of the
total Lagrangian follows from the nilpotency ${{\bf{\delta}}_B}^2=0$.
Thus the BRS symmetry is a symmetry of a gauge fixed Lagrangian,
so to say, a quantum gauge symmetry.
The corresponding conserved BRS charge reads as
\[
Q_B=\int d^3x\Big[
B^aD_0c^a-\partial_0B^a\cdot c^a+\displaystyle{i\over2}g
\partial_0 \bar c^a\cdot(c\times c)^a
\Big]
\]
where $(F\times G)^a=f_{abc}F^b G^c$. Thus one defines
physical space ${\cal V}_{phys}=\{|phys\rangle\}$ as
\[
Q_B |phys\rangle=0.
\]
It is to be noted that both ghost and anti-ghost fields are considered
as hermitian fermi fields so that the derived Hamiltonian should be
hermitian. This assignment is important for derivation of the unitarity,
but due to this assignment, there necessarily involves an indefinite
metric in a space of ghost and anti-ghost fields.
\item The BRS algebra is given by BRS charge, $Q_B$, and FP 
ghost charge, $Q_c$, as
\[
Q_B^2=0,\ \ \ [iQ_c,Q_B]=Q_B,\ \ \ [Q_c,Q_c]=0.
\]
Since these $Q_B, Q_c$ are commuting with other conserved charge,
all asymptotic one particle states can be classified by irreducible
representations. Due to the nilpotency of BRS charge, $Q_B$,
there only exist
BRS singles and doublets. From hermiticity of ghosts, $Q_c$ is defined
as a generator
of scale transformation of FP ghosts. It is to be noted
that while $Q_c$ being hermite,
FP ghost number is counted by $N_{FP}=iQ_c$, and
that among FP ghost number eigenstates, 
$\langle M|N\rangle \ne 0$ only if $M=-N$. Due to this metric
structure, BRS doublets always appear in pair of opposite sign
FP ghost numbers, and this pair
is called a BRS 4-tet (quartet).
Under the assumption that {\bf BRS singlets have positive metric},
it is proved that ${\bf {\cal V}_{phys}}$ has positive semidefinite
in such a way that {\bf BRS 4-tet
particles appear only in zero norm}.

\item From the Ward-Takahashi identities,
\begin{eqnarray}
F.T.[\langle 0|T D_{\mu}c^a(x)\bar c^b(y)|0\rangle
&=&-i \langle 0|T A^a_{\mu}(x)B^b(y)|0\rangle]\nonumber\\
&=&i \delta^{ab}\displaystyle{p_{\mu}\over p^2},\nonumber\\
\end{eqnarray}
where $F.T.$ implies the Fourier transformation,
it follows that Heisenberg operators,
$D_{\mu}c^a,\bar c^a, A^a_\mu, B^a$ necessarily have massless
asymptotic fields when $x_0\to \pm \infty$,
\[
A^a_{\mu}(x)\to \partial_{\mu}\chi^a(x)+\cdots,
\]
\[
B^a(x)\to \beta^a(x)+\cdots,
\]
\[
D_{\mu}c^a(x)\to\partial_{\mu}\gamma^a(x)+\cdots,
\]
\[
\bar c^a(x)\to \bar \gamma^a(x)+\cdots.
\]
One finds from the BRS transformation
that for each colour $a$, a set of the above massless asymptotic fields form a 
BRS 4-tet (quartet) such that
\[
[iQ_B,
\chi^a(x)
]=\gamma^a(x)
,\ \ \ 
\{iQ_B,\bar \gamma^a(x)\}
=i\beta(x)
\]
\[
\{iQ_B,
\gamma^a(x)
\}=0
,\ \ \ 
[iQ_B,
\beta^a(x)
]=0
\]
and
\[
\{\gamma^a(x),\bar \gamma^b(y)\}=
-i[\gamma^a(x),\bar \beta^b(y)]=-\delta^{ab}D(x-y)
\]
\item With respect to symmetry breaking in general, 
the following statements are
equivalent;
\begin{enumerate}
\item Charge of symmetry, $Q=\int d^3x j_0$, is well-defined.
\item Symmetry is not broken, $Q|0 \rangle =0$
\item There exist no massless one particle states in $j_{\mu}$ spectrum,
\[
\langle 0|j_{\mu}(x)|p^2=0\rangle=0
\]
\end{enumerate}

\item The Noether current corresponding to the conservation of the colour 
symmetry is $gJ^a_\mu={\partial ^\nu}{F^a_{\mu\nu}}+\{Q_B, D_\mu \bar c\}$,
where its ambiguity by divergence of antisymmetric tensor should be understood,
and this ambiguity is utilised so that massless contribution may be eliminated
for the charge, $Q^a$, to be well defined.

\item Denoting $g(A_{\mu}\times \bar c)^a\to u^a_b\partial_{\mu}\bar \gamma^b$,
and then $D_{\mu}\bar c^a\to (1+u)^a_b\partial_{\mu}\bar \gamma^b$,
one obtains that
\[
F.T.\langle 0|TD_\mu c^a(x)g(A_\nu\times \bar c)^b(y)|0\rangle
=(g_{\mu \nu}-{p_\mu p_\nu \over p^2})u^a_b(p^2),
\]
provided $A_{\mu}$ has a vanishing expectation value.
The current $\{ Q_B,D_\mu \bar c\}$ contains the massless component,
$(1+u)^a_b\partial_\mu \beta^b(x)$.
We can modify the Noether current for colour charge $Q^a$ such that
\[
gJ'^a_\mu=gJ_\mu-{\partial ^\nu}{F^a_{\mu\nu}}=\{Q_B, D_\mu \bar c\}.
\]
In the case of ${\bf 1+u=0}$, massless component in $gJ'_0$ is vanishing and
the colour charge 
\begin{equation}
Q^a=\int d^3x \{Q_B, g^{-1}D_0\bar c^a(x)\}
\label{kg}
\end{equation}
becomes {\bf well defined}.
\item
The physical state condition  $Q_B {\cal V}_{phys}=0$ together with the 
equation (\ref{kg}) implies that all BRS singlet one particle states
 $|f\rangle \in {\cal V}_{phys}$ are colour singlet states.
This statement implies that all coloured particles in ${\cal V}_{phys}$
belong to BRS 4-tet and have zero norm. This is the {\bf colour
confinement}.
\item
 In the course of their derivation, they assume Lorentz invariance and 
that the colour symmetry is not broken.
\item They also proved that if the vector massless asymptotic field
is missing in a channel $a$, and if the channel $a$ belongs to
the image of $1+u$ then the massless 4-tet in $j^a_{\mu}$
can not be cancelled, and the colour symmetry with charge  $Q^a$, 
is spontaneously broken. (Inverse Higgs mechanism theorem)
\end{enumerate}

The corresponding Euclidian expression
 is as follows,
\[
\int e^{-ip(x-y)}\langle  D_\mu c^a_x g(A_\nu\times \bar c)^b_y\rangle d^4x=
(\delta_{\mu\nu}-{p_\mu p_\nu\over p^2})u^{ab}(p^2),
\]
which 
can be calculated by
\begin{equation}
\displaystyle{1\over V}
\sum_{x,y} e^{-ip(x-y)}\left\langle  {\rm tr}\left({\lambda^a}^{\dag}
D_\mu \displaystyle{1\over \partial D}[-A_\nu\lambda^b] \right)_{xy}
\right\rangle=
(\delta_{\mu\nu}-{p_\mu p_\nu\over p^2})u^{ab}(p^2),
\label{eqq}
\end{equation}
where $\lambda^a$ is a normalized antihermitian basis of Lie algebra, $V$ a 
lattice volume,
and
the ghost propagator is given by
\begin{equation}
\langle c^a_x\bar c^b_y\rangle
=\left\langle{\rm tr}\left({\lambda^a}^{\dag}
\displaystyle{1\over \partial D}\lambda^b \right)_{xy}\right\rangle.
\label{eqq1}
\end{equation}

\subsection{Zwanziger's theory}
 The {\bf fundamental modular region} $\Lambda$ is specified by the
absolute minimum along the gauge orbits in the {\bf Gribov region} $\Omega$.
\begin{equation}
\Lambda=\{A|\|A\|^2={\rm Min}_g\|A^g\|^2\}, \qquad
\Lambda\subset \Omega=\{A|-\partial { D}\ge 0\ ,\ \partial A=0\}\ \ .
\label{NORM9}
\end{equation}

 Zwanziger relaxes the periodicity restriction on the gauge transformation
$g$, and imposes larger periods than the original. 
Then some two points in the fundamental modular region $\Lambda$
may be bridged to be Gribov copies of each other, and one of them is not
the absolute minimum of the minimizing function along the gauge orbit anymore.
Surviving points as the absolute minimum consist of {\bf core region} 
$\Xi$ ( $\Xi\subset\Lambda$).
In the so defined core region $\Xi$, a horizon function $H(U)$ given below is negative.
 
 The Horizon function is defined as follows.
Let two point tensor ${G_{\mu\nu xy}}^{ab}$ be
\begin{equation}
{G_{\mu\nu xy}}^{ab}
=
 {\rm tr}\left({\lambda^a}^{\dag}
D_\mu \displaystyle{1\over -\partial D}(-D_\nu)\lambda^b\right)_{xy}.
\label{eqq3}
\end{equation}
Then $H(U)$ is given as
\begin{equation}
H(U)=\sum_{x,y,a}{G_{\mu\mu xy}}^{aa}-(N^2-1)E(U)
\end{equation}
where $E(U)$ reads as follows;
\enmb
\item in $U$-linear version,
$E(U)=\sum_l{1\over N}{\rm Re}\ {\rm tr} U_l$,
\item in $A=\log U$ version,
$E(U)=\dis{{1\over N^2-1}\sum_{l,a}}{\rm tr}\left ({\lambda^a}^\dag S({\cal A}_l)\lambda^a\right)$,

where ${\cal A}_l=adj_{A_l}$, and 
$S(x)=\dis{{x/2\over {\rm th}(x/2)}}$.
\enme
Let us define an average tensor $G_{\mu\nu xy}$ be
$G_{\mu\nu xy}\delta^{ab}=\langle G_{\mu\nu xy}^{ab}\rangle,$
provided color symmetry is not broken.
One sees that a Fourier transform of the average tensor,
\[
G_{\mu\nu}(p)=\displaystyle{1\over V}
\sum_{x,y} e^{-ip (x-y)}G_{\mu\nu xy}
\]
takes a form
\begin{equation}
G_{\mu\nu}(p)\delta^{ab}=
\left(\dis{e\over d}\right)\displaystyle{p_\mu p_\nu\over p^2}\delta^{ab}
-\left(\delta_{\mu\nu}-\displaystyle{p_\mu p_\nu\over p^2}
\right)u^{ab},
\label{GMN}
\end{equation}
where
$e=\langle E(U)\rangle/V$,
and that it is related with the horizon function as
\begin{equation}
\displaystyle{\langle H(U)\rangle\over V}
=(N^2-1)\left[\lim_{p\to 0}
G_{\mu\mu}(p)-e\right].
\label{INFZW}
\end{equation}

 He defined the {\bf augmented core region} $\Psi=\{U : H(U)\le 0
\}\cap \Omega$ ( $\Xi\subset\Psi\subset \Omega$ ). $\Psi$ and $\Lambda$ are 
qualitatively similar, and he defined the partition function $Z_{\Psi}$
in the path integral in use of the corresponding Landau gauge function 
$f_{\Psi}(U)$, and concluded\cite{Zw} in the infinite volume limit that
$\lim_{V\to \infty}\displaystyle{\langle H(U)\rangle/ V}=0$. 
 Putting Kugo-Ojima parameter as
$u^{ab}(0)=-\delta^{ab}c,$
one finds from (\ref{GMN}), (\ref{INFZW}), that
\[
\left(\dis{e\over d}\right)+(d-1)c-e=(d-1)\left(c-\dis{e\over d}\right)=0,
\]
which is called horizon condition.
Since we can measure $c$ and $e$ by the lattice simulation, we can check
to what extent Zwanziger's horizon condition holds in
our simulation. With respect to the value $e/d$, note that the
classical vacuum is characterized by $e/d=1$.
\section{The numerical results}

\subsection{The algorithms of lattice Landau gauge fixing}

In the definition $A=\log U$, our Landau gauge fixing algorithm is as
follows. We define the gauge field\cite{NF}
on links as an element of $SU(3)$ Lie algebra as,
\begin{equation}
e^{A_{x,\mu}}=U_{x,\mu}\ \ ,\ \ \ {\rm where}\ \ A_{x,\mu}^{\dag}=-A_{x,\mu}.
\label{DFGAUGE}
\end{equation}
We perform the gauge transformation as
$e^{A^g_{x,\mu}}=g_x^\dagger  e^{A_{x,\mu}} g_{x+\mu}$
and define
$\displaystyle |\partial A|=Max_{x,\mu,a} |\partial A^a_{x}| ,
\|\partial A \|^2={1\over V(N^2-1)}\sum tr \partial A_{x}^\dagger \partial A_{x}$.
The Landau gauge is realized by minimizing $\|A^g\|^2$ via a gauge
transformation $g^\dagger U g$, where $g=e^\epsilon$.
In order to obtain $\epsilon$, we switch the following two methods, depending
on the current value of $|\partial A|$ in comparison to some critical 
parameter $|\partial A|_{cr}$.  

\begin{enumerate}
\item When $|\partial A|>|\partial A|_{cr}$,
$\epsilon_x={\eta'\over \|\partial A\|}\partial A_x$
with suitable parameter $1<\eta'<2.2$ 

\item
When $|\partial A|<|\partial A|_{cr}$,
 $\epsilon=(-\partial_\mu D_\mu(A))^{-1}\eta\partial A $
where $1<\eta<2$ is a parameter.
\end{enumerate}

The restriction to the fundamental modular region is not always achieved. 
But, we observed that the obtained norm $\|A\|$ is larger or smaller than that 
obtained after the smeared gauge fixing\cite{HdF} within 1\% accuracy.

In case of the $U$-linear definition of gauge field $A$, we perform the
site-local exact algorithm\cite{dFG}, with suitable over-relaxation parameter,
$\eta=1.6$, starting from gauge fixed configurations of $A=\log U$.
It is found that the exact over-relaxation $g=W^{1.6}$ is faster
than the stochastic over-relaxation, where $W$ is the site-local
exact solution obtained by solving the nonlinear equation
for finding the best fit gauge transformation on even-(odd-)site.

\subsection{The Kugo-Ojima two-point function and the ghost propagator}

 The FP operator is
\begin{equation}
{\cal M}[U]=-(\partial\cdot D(A))=-(D(A)\cdot\partial),
\end{equation}
and we define $Ad(A_\mu)$ by putting $D_\mu(A)=\partial_\mu+Ad(A_\mu)$.
The inverse, ${\cal M}^{-1}[U]=(M_0-M_1[U])^{-1}$, is calculated
 perturbatively by using the Green function of the  Poisson equation 
$M_0^{-1}=(-\partial^2)^{-1}$  and $M_1=\partial_\mu Ad (A_\mu)$
, as
\begin{equation}
{\cal M}^{-1}=M_0^{-1}+\sum_{k=0}^{N_{end}}(M_0^{-1}M_1)^kM_0^{-1}.
\end{equation}

 The ghost propagator (\ref{eqq1}) is infrared divergent and its singularity can be
parameterized as $\displaystyle p^{-2(1+\alpha)}$, where $\displaystyle p^2=\sum_{k\mu} (4 \sin^2{ \pi k_\mu\over L})$, $(-L/2 <k_\mu\le L/2)$. 
It depends on
 $\beta$ slightly, but its finite-size effect is small\cite{SS}.
These qualitative features are in agreement with the 
analysis of the Dyson-Schwinger equation\cite{SHA}.

\begin{figure}[htb]
\begin{minipage}[b]{0.47\linewidth}
\begin{center}
\epsfysize=100pt\epsfbox{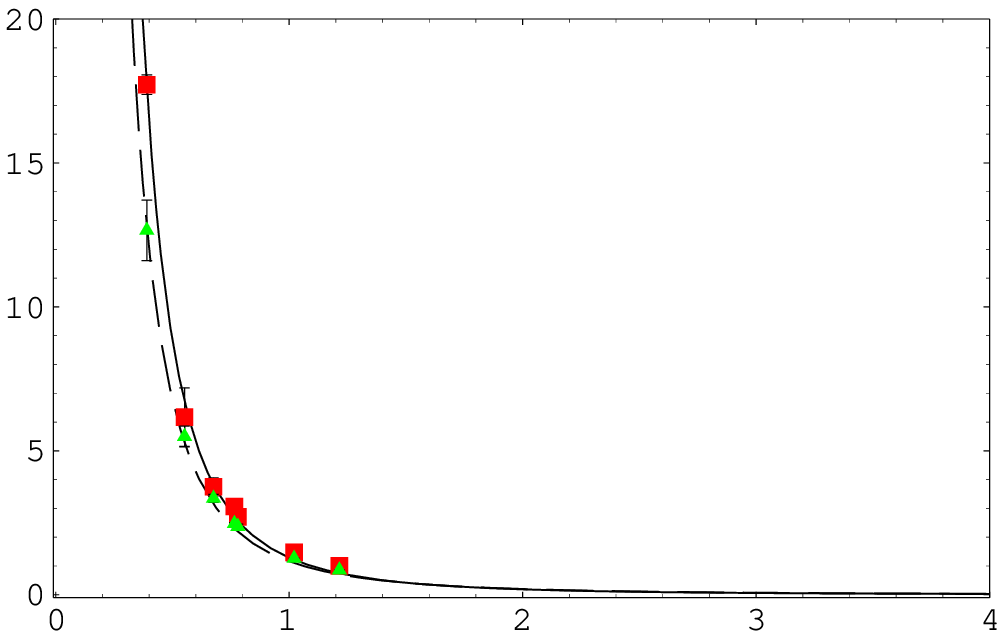}
\caption{The ghost propagator as function of the lattice momentum. 
The data are $\beta=5.5$(box) and $6.0$(triangle), $16^4$. The fitted curve is $1.287/p^{2.779}$ for $\beta=5.5$ and $1.162/ p^{2.545}$ for $\beta=6.0$ (dashed).}
\end{center}
\end{minipage}
\hfil
\begin{minipage}[b]{0.47\linewidth}
\begin{center}
\epsfysize=100pt\epsfbox{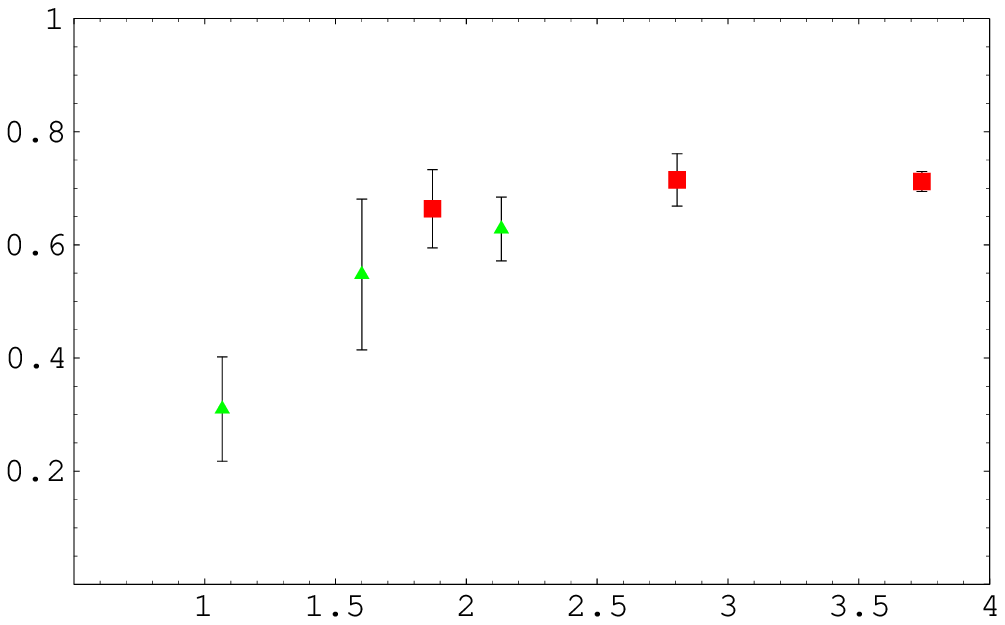}
\caption{The Kugo-Ojima parameter $|u^a_a|$ as the function of the spatial 
extent of the lattice $a L (fm)$.  The data are $\beta=6$ (triangle), and 
$\beta=5.5$ (box) $8^4, 12^4, 16^4$ from left to right, respectively.}
\end{center}
\end{minipage}
\end{figure}

We  measured the lattice version of  $|u^a_b(0)|$ on $8^3\times 16, 8^4$ 
$12^4$ and $16^4$ for $\beta=5.5$ and $6$. 
When $\beta=6$ and the lattice size is small, the Polyakov loop distribution 
deviates from the uniform distribution. In this case, we perform the 
${\Vec Z}_3$ rotation by multiplying the global phase $e^{\pm {2\pi\over 3}i}$
such that the distribution concentrate around one angle, before we measure
the Kugo-Ojima two-point function.

We obtained that $u^a_b(0)$ is consistent with $-c\delta^a_b,\ c=0.7$ in
$SU(3)$ quenched simulation, $\beta =5.5$, on $8^4, 12^4$ and $16^4$.

\subsection{The gluon propagator }

 The gluon propagator is infrared finite. We parameterized the zero-temperature
lattice data using the Stingl's Factorised Denominator Rational Approximant
(FDRA) method. The effective mass of the gluon in the analysis of $8^3\times 16$ is found to be about $600 MeV$.

The infrared finiteness is in accordance with the Kugo-Ojima color confinement
mechanism. 
 As stated in the their inverse Higgs mechanism theorem, 
if we have no massless vector poles in all channels of the gauge field,
$A^a_{\mu}$, and if the color symmetry is not broken at all,
it follows that $1+u=0$.\cite{Iz}.

\section{Discussion and conclusions}

We performed the first test of the Kugo-Ojima color confinement
criterion in lattice Landau gauge. We observed that the $8^4, 12^4$ and $16^4$
 lattice data. The data of $\beta=5.5$ indicates that $u=-0.7$, and those
of $\beta=6.0$ are smaller by about 10\%.

 In the Zwanziger's theory, the two-point function $G_{\mu\nu}(k^2)$
can be expressed in terms of the Kugo-Ojima two-point function as (\ref{GMN}).
Zwanziger's horizon condition\cite{Zw} in the infinite volume limit reads as
$G_{\mu\mu}(0)=e=\langle {E(U)\over  V}\rangle$.
In terms of the Kugo-Ojima parameter $c$, the left hand side can be written as 
$(e/4)+3c$, and the horizon condition is that $c=e/d$. 
In the table below $e_1$
and $e_2$ stand for $e$ in our $16^4$ lattice simulation of the first
and the second option of the gauge fields, respectively.

If the gauge fixing could be performed so that it brings the configuration into
the core region or the augmented core region and if the infinite volume
limit is considered somehow, then the legitimate check of the horizon
condition could be done. The core gauge fixing is, however, difficult, and
even impossible in general, which implies that the core gauge is literarily
 not the gauge. A configuration of the core region of period $L$, belongs
to a fundamental modular region of larger period $NL$ as well as to
 a fundamental modular region of period $L$ by
definition\cite{Zw}.
Such a configuration is particular one in the fundamental modular region of
period $L$, to which a generic configuration in the fundamental modular
region of period $L$ cannot be gauge transformed even by a relaxed gauge
transformation of larger period. A relaxed gauge transformation with
larger period $NL$ unique up to global gauge transformation 
can bring the generic configuration above to
the fundamental modular region of period $NL$, but at the same time
 breaks the periodicity of $L$ in general. Thus restriction
to core region is neihter an argument of 'gauge' nor an
argument of trace disappearance of periodicity. It is highly
dynamical hypothesis that the core region and the fundamental modular region
give the same limiting correlation fucntions, i.e., dynamics, in the
infinite volume limit. 
Thus we take a standpoint that the horizon condition derived from
Zwanziger's restriction to the core region is simply tested by the dynamics
of the fundamental modular region, and
give the direct results in the table although obviously not in the
infinite volume limit.

\begin{table}[htb]
\caption{$\beta$ dependence of the Kugo-Ojima parameter $c$, 
 the tensor $G_{\mu\mu}/4$, trace $e$ divided by the dimension $d$.
The suffix 1 corresponds to the $u$-linear and 2 corresponds to  the $\log U$
definition. Data are those of $16^4$, except  $\beta=5.5$ $U$-linear data,
which are those of $8^4$. } 
\vskip 0.5 true cm
\begin{tabular*}{\textwidth}{@{}l@{\extracolsep{\fill}}c|ccc|cccc}
    & $\beta$    & $c_1$  & $e_1/d$  & $G_{\mu\mu 1}(0)/d$&$c_2$& $e_2/d$& $G_{\mu\mu 2}(0)/d$ &\\
\hline
 &5.5  & 0.570(58)& 0.780(3) & 0.622(45) &0.712(18)  &0.657(1) &0.698(14) \\
 &6.0   &0.576(79)& 0.860(1) & 0.647(57) & 0.628(94)  & 0.693(1)  & 0.644(70) \\
\hline
\end{tabular*}
\end{table}

Simulation data show in general that when $\beta$ becomes larger, $e$
becomes larger, while $c$ has an opposite tendency. This fact itself
does not necessarily disprove the horizon condition, but
our data of $c$ which is calculated in the $A=\log U$ version
 already
gives the zero-intersection of $G_{\mu\mu}/d-e_2/d$ in the increase of $\beta$
from $5.5$ to $6$.

This work was supported by KEK Supercomputer Project(Project No.00-57), and by JSPS, Grant-in-aid for Scientific Research(C) (No.11640251).

\end{document}